\documentclass[10pt]{iopart}
\usepackage[latin1
]{inputenc}
\usepackage{graphicx,ifthen,iopams,setstack,amssymb,rotating}
\jl{1}
\eqnobysec
\begin{document}
\title{Statistically interacting quasiparticles in Ising chains} 
\author{
Ping Lu$^{1}$, 
Jared Vanasse$^{1}$, 
Christopher Piecuch$^{1}$,  
Michael Karbach$^{1,2}$, 
and Gerhard M{\"u}ller$^1$
}
\address{
$^1$Department of Physics, University of Rhode Island,
  Kingston RI 02881, USA \\
$^2$Fachbereich Physik, Bergische Universit{\"a}t Wuppertal, 
  42097 Wuppertal, Germany \\
}

\begin{abstract}
  The exclusion statistics of two complementary sets of quasiparticles,
  generated from opposite ends of the spectrum, are identified for Ising chains
  with spin $s=1/2,1$. In the $s=1/2$ case the two sets are antiferromagnetic
  domain walls (solitons) and ferromagnetic domains (strings).  In the $s=1$
  case they are soliton pairs and nested strings, respectively.  The Ising
  model is equivalent to a system of two species of solitons for $s=1/2$ and to
  a system of six species of soliton pairs for $s=1$.  Solitons exist on single
  bonds but soliton pairs may be spread across many bonds. The thermodynamics
  of a system of domains spanning up to $M$ lattice sites is amenable to exact
  analysis and shown to become equivalent, in the limit $M\to\infty$, to the
  thermodynamics of the $s=1/2$ Ising chain. A relation is presented between
  the solitons in the Ising limit and the spinons in the $XX$ limit of the
  $s=1/2$ $XXZ$ chain.
\end{abstract}
%
\section{Introduction}\label{sec:intro}
%

Ising chains are among the simplest systems of interacting degrees of freedom
and have been thoroughly studied in a wide variety of circumstances
including the presence of transverse fields, time-dependent fields,
inhomogeneities in field or coupling etc. Is there anything of substance left
that we can still learn from the Ising model in one dimension
\cite{Isin25} with homogeneous coupling? This paper presents a case (by no
means the only one \cite{ADR04, DH05}) for an affirmative answer.

The Hamiltonian of the spin-$s$ Ising model for $s=1/2,1,3/2,\ldots$ on a
periodic chain of $N$ sites reads
\begin{equation}
  \label{eq:10}
  \mathcal{H}_s=\sum_{n=1}^N\left(JS_n^zS_{n+1}^z +hS_n^z\right),\quad 
S_n^z=s,s-1,\ldots,-s,
\end{equation}
where the exchange coupling is antiferromagnetic (ferromagnetic) for $J>0$
$(J<0)$ and $h$ is a magnetic field.  This model system has simple product
eigenstates, a dispersionless spectrum, and no intrinsic dynamics to speak of.
Its thermodynamics, derived via transfer matrix \cite{KW41,
  STK67}, is predictably simple.

One interesting aspect of $\mathcal{H}_s$ that promises usefulness in a wider
context is the quasiparticle composition of the product eigenstates as will be
demonstrated. The entire spectrum of $\mathcal{H}_s$ can be systematically
generated from opposite ends by different sets of quasiparticles with exotic
exclusion statistics. The nature of these quasiparticles strongly varies with
$s$ but a systematics in their make-up is recognizable.

In $\mathcal{H}_{1/2}$ we consider antiferromagnetic domain walls (solitons
with spin $\pm1/2$ and fractional exclusion statistics) for $J>0$ or with
ferromagnetic domains (strings of flipped spins with integer-valued exclusion
statistics) for $J<0$. The corresponding quasiparticles in $\mathcal{H}_1$ turn
out to be soliton pairs with spin $0,\pm1$ (for $J>0$) and nested strings (for
$J<0$), both with unusual exclusion statistics.

We use the concept of statistically interacting quasiparticles to show that the
thermodynamics of $\mathcal{H}_s$ is equivalent to that of a gas of solitons
(for $s=1/2$) or soliton pairs (for $s=1$). The same framework is shown to work
also for the thermodynamics of string particles.  It is expected that these
particles, whose detailed exclusion statistics is worked out here, are still
relevant in integrable spin chain models away from their Ising limit.  The
particles identified here then become objects of a coordinate Bethe ansatz
\cite{KBI93, Suth04, EFG+05} applied to those models.

We first review the concept of statistical interaction and its use in
thermodynamics (Sec.~\ref{sec:statint}). Then we introduce the soliton
particles for $\mathcal{H}_{1/2}$, describe their exclusion statistics, and
determine the thermodynamics in a magnetic field from a soliton perspective
(Sec.~\ref{sec:isoh}). Next we introduce the six species of soliton-pair
particles that govern the spectrum of $\mathcal{H}_1$ and work out their
thermodynamics in zero magnetic field (Sec.~\ref{sec:isone}). Then we present
the combinatorics for the statistical interaction of a system of strings in
$\mathcal{H}_{1/2}$ and of nested strings in $\mathcal{H}_1$. We proceed by
calculating the thermodynamics of a system of strings of restricted size and
recover the Ising thermodynamics when that restriction is lifted
(Sec.~\ref{sec:stri}). Finally, we assess the progress reported here and
discuss possible extensions and comparisons (Sec.~\ref{sec:outlook}) including
a relation between solitons and spinons, both with semionic statistics
(\ref{sec:nestri}).


%
\section{Statistical interaction}\label{sec:statint}
%

Quasiparticles in solid matter are not restricted to be either bosons or
fermions. In integrable quantum many-body model systems \cite{KBI93, Suth04,
  EFG+05} quasiparticles with infinite lifetimes and unusual exclusion
statistics have indeed been identified.  The generalized Pauli principle as
introduced by Haldane \cite{Hald91a} expresses how the number of states
available to one particle is affected by the presence of other particles:
\begin{equation}\label{eq:core}
  \Delta d_i \doteq -\sum_j g_{ij}\Delta N_j.
\end{equation}
The indices $i,j$ refer to distinct particle species. The $g_{ij}$ are {\em
  statistical interaction coefficients}. For bosons we have $g_{ij}=0$ and for
fermions $g_{ij}=\delta_{ij}$.  Upon integration Eq.~(\ref{eq:core}) becomes
\begin{equation}\label{eq:d1p}
  d_i = A_i -\sum_j g_{ij}(N_j-\delta_{ij}),
\end{equation}
where the $A_i$ are {\em statistical capacity constants}. The number of
many-body states containing $\{N_i\}$ particles of the various species is then
determined by the multiplicity expression
\begin{equation}\label{eq:prodWa}
  W(\{N_i\})=\prod_i\left( \begin{tabular}{c} $d_i+N_i-1$ \\ $N_i$ 
\end{tabular}\right).
\end{equation}
We shall determine the ingredients $A_i, g_{ij}$ to (\ref{eq:d1p}) for two
species of solitons or $N$ species of strings, all pertaining to
$\mathcal{H}_{1/2}$, and for six species of soliton pairs or $N(N+1)/2$ species
of nested strings pertaining to $\mathcal{H}_1$.

The thermodynamic properties of a macroscopic system of statistically
interacting particles are amenable to a rigorous analysis as shown by Wu
\cite{Wu94}. For given sets of 1-particle energies $\epsilon_i$, statistical
interaction coefficients $g_{ij}$, and statistical capacity constants $A_i$,
the grand partition function is
\begin{equation}
  \label{eq:71bb}
  Z=\prod_i\left[\frac{1+w_i}{w_i}\right]^{A_i},
\end{equation}
where the quantities $w_i$ are determined by the nonlinear algebraic
equations
\begin{equation}
  \label{eq:54bb}
  \frac{\epsilon_i-\mu}{k_BT} = 
\ln(1+w_i)-\sum_jg_{ji}\ln\left(\frac{1+w_j}{w_j}\right).
\end{equation}
The temperature $T$ and the chemical potential $\mu$ are the control variables.
Additional control variables such as external fields may come into play as part
of the energies $\epsilon_i$. The average numbers of particles, $\langle
N_i\rangle$, of each species are related to the $w_i$ by the linear equations
\begin{equation}
  \label{eq:16appb}
  w_i\langle N_i\rangle+\sum_jg_{ij}\langle N_j\rangle=A_i.
\end{equation}
We shall apply this method of exact analysis to the solitons and the strings in
the context of $\mathcal{H}_{1/2}$ and to soliton pairs in the context of 
$\mathcal{H}_1$.


%
\section{Solitons}\label{sec:isoh}
%

Here we consider $\mathcal{H}_{1/2}$ with $J>0$ and $h>0$ for even or odd $N$.
The task at hand has a combinatorial part and a statistical mechanical part. We
first relate the Ising spectrum to soliton configurations, then we undertake a
thermodynamic analysis of the soliton system, using the methodology outlined in
Sec.~\ref{sec:statint}.

\subsection{Combinatorics of solitons}\label{sec:solconspec}
Among the four distinct bonds in the general product state
$|\sigma_1\sigma_2\cdots\sigma_N\rangle$ (see Table~\ref{tab:akscs1h}), the
bonds $\uparrow\uparrow$, $\downarrow\downarrow$ represent solitons with spin
$+1/2$, $-1/2$, respectively, and $\uparrow\downarrow$, $\downarrow\uparrow$
are vacuum bonds.  Close-packed solitons with like spin orientation reside on
successive bonds (e.g.  $\uparrow\uparrow\uparrow$), whereas close-packed
solitons with opposite spin orientation are separated by one vacuum bond (e.g.
$\uparrow\uparrow\downarrow\downarrow$). More generally, the number of vacuum
bonds between nearest-neighbor solitons with like (opposite) spin orientation
is even (odd). Solitons only interact statistically. The energy of a soliton is
unaffected by the presence of other solitons.

\begin{table}[ht]
  \caption{Distinct bonds in $\mathcal{H}_{{1/2}}$, their  
    soliton content, and their contribution to the energy of the product
    eigenstate (relative to the soliton vacuum).} \label{tab:akscs1h}  
\begin{center}
\begin{tabular}{r|cccc} \rule[-2mm]{0mm}{5mm}
$\mathrm{bond}$ & $\uparrow\uparrow$ & $\downarrow\downarrow$ & 
$\uparrow\downarrow$ & $\downarrow\uparrow$ \\ \hline \rule[-2mm]{0mm}{6mm}
$N_+$ & $1$ & $0$ & $0$ & $0$ \\ \rule[-2mm]{0mm}{5mm}
$N_-$ & $0$ & $1$ & $0$ & $0$ \\ \hline \rule[-2mm]{0mm}{6mm}
$\Delta E$ & $\frac{J+h}{2}$ & $\frac{J-h}{2}$ & $0$ & $0$
\end{tabular}
\end{center}
\end{table} 

The two soliton vacuum states,
$|\uparrow\downarrow\cdots\uparrow\downarrow\rangle$ and
$|\downarrow\uparrow\cdots\downarrow\uparrow\rangle$, represent the lowest
energy level for even $N$. The lowest level for odd $N$ is $2N$-fold degenerate
and contains one soliton. The soliton content of an Ising eigenstate is
specified either by the numbers $N_\pm$ of spin-up/down solitons or,
alternatively, by the total number of solitons and the magnetisation:
\begin{equation}
  \label{eq:12plj1}
  N_A=N_+ +N_-,\quad M_z=\frac{1}{2}(N_+ -N_-).
\end{equation}
The energy level of all states with $N_A$ solitons and magnetisation $M_z$ is
\begin{equation}
  \label{eq:3plj}
 E(N_A,M_z)-E_0 = \frac{1}{2}N_AJ+hM_z,  
\end{equation}
where $E_0=-NJ/4$ is the energy of the soliton vacuum.

How many Ising eigenstates exist for given $N_+$ and $N_-$ (or $N_A$ and
$M_z$)? The solution of this combinatorial problem is the following multiplicity
expression constructed from extensive tabulated data such as sampled in
Table~\ref{tab:wadat89}:
\begin{equation}
  \label{eq:7clj}
  W_A(N_+,N_-) = \frac{2N}{N-N_A}\;\prod_{\sigma=\pm}  
\left(
\begin{tabular}{c}
$d_\sigma +N_\sigma -1$ \\ $N_\sigma$
\end{tabular}
\right),
\end{equation}
\begin{equation}
  \label{eq:8clj}
  d_\sigma=\frac{1}{2}(N-1) -\frac{1}{2}\sum_{\sigma'}(N_{\sigma'}-\delta_{\sigma\sigma'}). 
\end{equation}
It is indeed compatible with the standard form (\ref{eq:prodWa}).  The range of
$N_A$ is $0,2,\cdots,N$ for even $N$ and $1,3,\cdots,N$ for odd $N$. $N_A=N$ is
only realized for the two states with $N_+=N_A$ or $N_-=N_A$.  This
multiplicity expression specifies the statistical interaction between soliton
particles.
\begin{table}[hb]
  \caption{Number of states, $W_A(N_+,N_-)$, with $N_A=N_+ +N_-$
solitons and magnetisation $M_z=\left(N_+ -N_-\right)/2$ for
$\mathcal{H}_{1/2}$ 
with $N=6$ (left) and $N=7$ (right).} \label{tab:wadat89} 
\begin{center}
\begin{tabular}{r|cccc|c}
$M_z\backslash N_A$ & 0 & 2 & 4 & 6 & \\ \hline
$3$~ & -- & -- & -- & 1 & 1 \\
$2$~ & -- & -- & 6 & -- & 6 \\
$1$~ & -- & 9 & 6 & -- & 15 \\
$0$~ & 2 & 12 & 6 & -- & 20 \\
$-1$~ & -- & 9 & 6 & -- & 15 \\
$-2$~ & -- & -- & 6 & -- & 6 \\
$-3$~ & -- & -- & -- & 1 & 1 \\ \hline
 & 2 & 30 & 30 & 2 & 64 \\ 
\end{tabular}\hspace*{10mm}
\begin{tabular}{r|cccc|c}
$M_z\backslash N_A$ & 1 & 3 & 5 & 7 & \\ \hline
$7/2$~ & -- & -- & -- & 1 & 1 \\
$5/2$~ & -- & -- & 7 & -- & 7 \\
$3/2$~ & -- & 14 & 7 & -- & 21 \\
$1/2$~ & 7 & 21 & 7 & -- & 35 \\
$-1/2$~ & 7 & 21 & 7 & -- & 35 \\
$-3/2$~ & -- & 14 & 7 & -- & 21 \\
$-5/2$~ & -- & -- & 7 & -- & 7 \\
$-7/2$~ & -- & -- & -- & 1 & 1 \\ \hline
 & 14 & 70 & 42 & 2 & 128 \\ 
\end{tabular}
\end{center}
\end{table} 


\subsection{Statistical mechanics of 
solitons}\label{sec:stamesoga} 

For the statistical mechanical analysis of $\mathcal{H}_{1/2}$ as a soliton gas
we use the statistical capacity constants $A_{\sigma}=(N-1)/2$ and the
statistical interaction coefficients $g_{\sigma\sigma'}=1/2$ extracted from
Eq.~(\ref{eq:8clj}), and the soliton energies $\epsilon_{\sigma}=(J+\sigma
h)/2$ from Table~\ref{tab:akscs1h}. We have to solve two coupled
nonlinear algebraic equations of the type (\ref{eq:54bb}): 
  \begin{equation}
    \label{eq:1}
    \frac{J\pm h}{2k_BT}= \ln(1+w_\pm) +\frac{1}{2}\ln\frac{w_\pm}{1+w_\pm}+
  \frac{1}{2}\ln\frac{w_{\mp}}{1+w_\mp}.
  \end{equation}
The solution,
\begin{equation}
  \label{eq:2}
  w_\pm= \frac{1}{2}\left[e^{\pm h/k_BT}-1+ \sqrt{(e^{\pm h/k_BT}-1)^2+4e^{(J\pm
        h)/k_BT}}\right],
\end{equation}
determines the grand partition function via (\ref{eq:71bb}) with the
asymptotic value $A_\pm\leadsto N/2$ for the capacity constants. The
result,
\begin{eqnarray}
   \label{eq:3}
\fl   Z= \left[e^{2K}\left(\cosh H+\sqrt{\sinh^2H+ e^{-4K}}\, 
     \right)\right]^N,\quad
   K\doteq-\frac{J}{4k_BT},\quad H\doteq-\frac{h}{2k_BT},
\end{eqnarray}
is in exact agreement with the well-known canonical partition function $Z_N$
obtained via transfer-matrix \cite{KW41}. In the relation $Z=Z_Ne^{NK}$,
the factor $e^{NK}$ accounts for the relative shift of energy scales used in
the two methods.

\begin{figure}[b!]
  \centering
  \includegraphics[width=67mm,angle=-90]{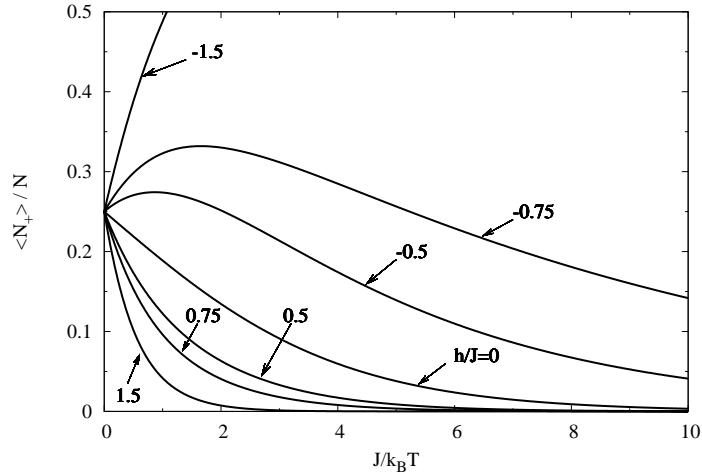}  
  \caption{Average number $\langle N_+\rangle/N$ of spin-up solitons
    per lattice bond versus $J/k_BT$ for several values of magnetic field.}
  \label{fig:qpis15a}
\end{figure}

For the average numbers of solitons, $\langle N_\pm\rangle$, we infer from
(\ref{eq:16appb}) the two coupled linear equations,
\begin{equation}
  \label{eq:4}
  \left(w_\pm +\frac{1}{2}\right)\langle N_\pm\rangle + 
\frac{1}{2}\langle N_\mp\rangle =\frac{N}{2},
\end{equation}
which have the solutions
\begin{equation}
  \label{eq:5}
  \hspace*{-24mm}\langle N_\pm\rangle =  
\frac{N}{2}\,\frac{e^{\pm H}\left[\sqrt{\sinh^2H+ e^{-4K}}\pm 
    \sinh{H}\right]}%
{\sinh^2H+e^{-4K} +\cosh{H}\sqrt{\sinh^2H+ e^{-4K}}}~ 
\stackrel{h\to0}{\longrightarrow} ~\frac{N/2}{e^{-2K}+1}.
\end{equation}
The dependence of $\langle N_{+}\rangle/N$ on $J/k_BT$ is shown in
Fig.~\ref{fig:qpis15a} for various values of $h/J$. 
All curves start from $\langle N_+\rangle/N=1/4$ in the high-$T$ limit. For
$h=0$ we have $\langle N_+\rangle=\langle N_-\rangle$; this curve has a
monotonically decreasing trend toward zero as $T\to0$.  For $h>0$ $(h<0)$ the
average number $\langle N_+\rangle$ of solitons with spin directed antiparallel
(parallel) to $h$ is more (less) rapidly suppressed as $T\to0$. For
sufficiently weak, negative fields, $0>h/J>-0.25$, the curve is still
monotonically decreasing.  For $-0.25>h/J>-1$, it acquires a smooth maximum at
finite, nonzero $T$. For $h/J<-1$ the curve is monotonically increasing toward
$\langle N_{+}\rangle/N$=1.  Here the ground state contains $N$ spin-polarized
solitons.


%
\section{Soliton pairs}\label{sec:isone}
%

Here we consider $\mathcal{H}_1$ with $J>0$ and $h=0$ for even or odd $N$.  The
generalization to $h\neq0$ is straightforward conceptually.  The independent
particles are now soliton pairs.

\subsection{Combinatorics of soliton pairs}\label{sec:solpaiconspec}

The nine different kinds of bonds are listed in Table~\ref{tab:akscs1}.  Each
bond can accommodate up to two solitons. The energy of a soliton is not the
same in all configurations.

\begin{table}[h!]
  \caption{Distinct bonds in $\mathcal{H}_1$, their soliton
    content, and their contribution to the energy of a product eigenstate
    (relative to the soliton vacuum).}\label{tab:akscs1} 
\begin{center}
\begin{tabular}{r|ccccccccc}
$\mathrm{bond}$ & $\uparrow\uparrow$ & $\circ\circ$ & 
$\downarrow\downarrow$ & $\uparrow\circ$ & $\circ\uparrow$ & 
$\downarrow\circ$ & $\circ\downarrow$ &
$\uparrow\downarrow$ & $\downarrow\uparrow$ \\ \hline \rule[-2mm]{0mm}{7mm}
$N_+$ & $2$ & $1$ & $0$ & $1$ & $1$ & $0$ & $0$ & $0$ & $0$\\
\rule[-2mm]{0mm}{5mm}
$N_-$ & $0$ & $1$ & $2$ & $0$ & $0$ & $1$ & $1$ & $0$ & $0$\\ \hline
\rule[-2mm]{0mm}{6mm}
$\Delta E/J$ & $2$ & $1$ & $2$ & $1$ & $1$ & $1$ & $1$ & $0$ & $0$\\
\end{tabular}
\end{center}
\end{table} 

Our search for the independent particles again starts from extensive tabulated
data for $W_A(N_+,N_-)$ such as sampled in Table~\ref{tab:wadat34}. Several
clues suggest that the independent particles are soliton pairs. For example,
the number of solitons is always even. Also, the number of states with
$N_{-}=0$ grows $\propto N^{N_+/2}$ for $N_+\ll N$ as opposed to
the growth $\propto N^{N_+}$ observed in $\mathcal{H}_{1/2}$

\begin{table}[tb]
  \caption{Number of states, $W_A(N_+,N_-)$, with $N_A=N_+ +N_-$
solitons and magnetisation $M_z=\left(N_+ -N_-\right)/2$ for $\mathcal{H}_1$
with $N=3$ (left) and $N=4$ (right).} \label{tab:wadat34} 
\begin{center}
\begin{tabular}{r|cccc|c}
$M_z\backslash N_A$ & 0 & 2 & 4 & 6 & \\ \hline
3~ &  -- & -- & -- & 1 & 1 \\
2~ &  -- & -- & 3 & -- & 3 \\
1~ &  -- & 3 & 3 & -- & 6 \\
0~ &  -- & 6 & -- & 1 & 7 \\
$-1$~ &  -- & 3 & 3 & -- & 6 \\
$-2$~ &  -- & -- & 3 & -- & 3 \\
$-3$~ &  -- & -- & -- & 1 & 1 \\\hline
 & 0 & 12 & 12 & 3 & 27 \\ 
\end{tabular}
\hspace*{5mm}
\begin{tabular}{r|ccccc|c}
$M_z\backslash N_A$ & 0 & 2 & 4 & 6 & 8 & \\ \hline
4~ & -- & -- & -- & -- & 1 & 1 \\
3~ & -- & -- & -- & 4 & -- & 4 \\
2~ & -- & -- & 6 & 4 & -- & 10 \\
1~ & -- & 4 & 8 & 4 & -- & 16 \\
0~ & 2 & -- & 16 & -- & 1 & 19 \\
$-1$~ & -- & 4 & 8 & 4 & -- & 16 \\
$-2$~ & -- & -- & 6 & 4 & -- & 10 \\
$-3$~ & -- & -- & -- & 4 & -- & 4 \\
$-4$~ & -- & -- & -- & -- & 1 & 1 \\ \hline
 & 2 & 8 & 48 & 20 & 3 & 81 \\ 
\end{tabular}
\end{center}
\end{table} 

The systematic examination of the data tables for $W_A(N_+,N_-)$ points to the
existence of six distinct species of soliton-pair particles, two groups of
three species with spin $(+1,0,-1)$. In the first group the paired solitons are
confined to the same bond. In the second group the paired solitons are
deconfined. They can be on bonds with any number of lattice sites between them.

Confined-soliton pairs with spin up (named $r_+$) are identified by any element
$\uparrow\uparrow$ in the product state. In like manner, confined-soliton pairs
with spin zero (down) are named $r_0$ $(r_-)$ and identified by elements
$\circ\circ$ $(\downarrow\downarrow)$ in the product state.  Deconfined-soliton
pairs with spin up (down) are named $q_+$ $(q_{-})$ and identified by any
element $\uparrow\circ\cdots\circ\uparrow$
$(\downarrow\circ\cdots\circ\downarrow)$ in the product state, where the
presence of $n=1,2,\ldots$ site variables $\circ$ between two site variables
$\uparrow$ $(\downarrow)$ indicate the presence of $n-1$ spin-zero
confined-soliton pairs $(r_0)$. Deconfined-soliton pairs with spin zero are
named $q_0$ and are identified by elements $\uparrow\circ\cdots\circ\downarrow$
or $\downarrow\circ\cdots\circ\uparrow$ in the product state.

A list of name, motif, and soliton content for all six species of soliton-pair
particles is shown in the top three rows of Table~\ref{tab:solpqp1}.  In some
instances, two close-packed particles share one lattice site (e.g.
$\circ\circ\circ$, $\uparrow\circ\uparrow\uparrow$,
$\uparrow\circ\downarrow\circ\downarrow$), in other instances, there is one
vacuum bond in between (e.g. $\uparrow\uparrow\downarrow\downarrow$,
$\uparrow\circ\uparrow\downarrow\circ\uparrow$).  The particle $r_0$ can only
exist inside one of the particles $q_+,q_0,q_-$. The former is instrumental to
the soliton deconfinement in the latter.

The six particles are thus naturally classified into three groups, the
confined-soliton pairs $r_+,r_-$, the deconfined-soliton pairs $q_+,q_0,q_-$,
and the spacer particle $r_0$ (deconfinement agent). In a vague QCD analogy,
solitons play the role of quarks, the soliton pairs $r_+,r_-,q_+,q_0,q_-$ the
role of mesons and baryons, and the spacer particle $r_0$ the role of gluon
with opposite action.

\begin{table}[htb]
  \caption{Specifications of particles in
    $\mathcal{H}_1$: confined-soliton pairs $(r_+,r_-)$,
    spacer particle $(r_{0})$, 
    deconfined-soliton pairs $(q_+,q_0,q_-)$; motif in product state;
    soliton content; index $m$ used in (\ref{eq:48aw6}); statistical capacity
    constants $A_m$; energies $\epsilon_m$.} \label{tab:solpqp1} 
\begin{center}
\begin{tabular}{r||cc|c|ccc} 
$\mathrm{particle}$ & $r_+$ & $r_-$ & $r_0$ & $q_+$ & $q_-$ & $q_0$ \\ \hline
\rule[-2mm]{0mm}{6mm}  
$\mathrm{motif}$ & $\uparrow\uparrow$ & $\downarrow\downarrow$ & $\circ\circ$ &
$\uparrow\circ\uparrow$ & $\downarrow\circ\downarrow$ &
$\uparrow\circ\downarrow$, $\downarrow\circ\uparrow$ \\ 
\rule[-2mm]{0mm}{5mm}   
$N_++N_-$ & $2+0$ & $0+2$ & $1+1$ & $2+0$ & $0+2$ & $1+1$ \\ \hline
\rule[-2mm]{0mm}{6mm}   
$m$ & $1$ & $2$ & $3$ & $4$ & $5$ & $6$\\ \rule[-2mm]{0mm}{5mm} 
$A_m$ & $\frac{N-1}{2}$ & $\frac{N-1}{2}$ & $0$ & $\frac{N}{2}-1$ &
$\frac{N}{2}-1$ & 
$N-2$ \\ \rule[-2mm]{0mm}{5mm} 
$\epsilon_m$ & $2J$ & $2J$ & $J$ & $2J$ & $2J$ & $2J$
\end{tabular}
\end{center}
\end{table} 

We have determined the multiplicity expression
\begin{equation}
  \label{eq:48aw6}
  W_6(\{X_m\})=\frac{2N}{N-N_\Sigma} \prod_{m=1}^6
\left( \begin{tabular}{c}
$d_m+X_m-1$ \\ $X_m$
\end{tabular} \right),
\end{equation}
\begin{equation}
  \label{eq:48w6}
  d_m=A_m-\sum_{m'}g_{mm'}(X_{m'}-\delta_{mm'}),
\end{equation}
\begin{equation}
  \label{eq:47w6}
  N_\Sigma=X_1+X_2+X_3 +2(X_4+X_5+X_6)\leq N,
\end{equation}
for product eigenstates containing $X_m$ soliton pairs of species
$m=1,\ldots,6$, where the index $m$ is defined in Table~\ref{tab:solpqp1}. It
confirms the independent status of the soliton-pair particles and contains the
specifications of their statistical interaction. The capacity constants $A_m$
and the particle energies $\epsilon_m$ are given in Table~\ref{tab:solpqp1},
and the interaction coefficients $g_{mm'}$ in Table~\ref{tab:gmmps1w6}.  Again
there exist restrictions and exceptions regarding the allowed configurations
$\{X_m\}$. We do not list them here because they have no bearing on the
statistical mechanical analysis.  The only model specifications needed are the
quantities $A_m, \epsilon_m, g_{mm'}$.

\begin{table}[htb] 
  \caption{Statistical interaction coefficients $g_{mm'}$ between soliton-pair
    particles as identified in 
    Table~\ref{tab:solpqp1}.}\label{tab:gmmps1w6}     
\begin{center} 
\begin{tabular}{c|rrrrrr} 
$g_{mm'}$ & $1$ & $2$ & $3$ & $4$ & $5$ & $6$\\ \hline \rule[-2mm]{0mm}{6mm} 
$1$ & $\frac{1}{2}$ & $\frac{1}{2}$ & $\frac{1}{2}$ & $0$ & $1$ &
$\frac{1}{2}$\\ \rule[-2mm]{0mm}{5mm}  
$2$ & $\frac{1}{2}$ & $\frac{1}{2}$ & $\frac{1}{2}$ & $1$ & $0$ & $\frac{1}{2}$
\\ \rule[-2mm]{0mm}{5mm}  
$3$ & $0$ & $0$ & $0$ & $-1$ & $-1$ & $-1$ \\ \rule[-2mm]{0mm}{5mm}  
$4$ & $\frac{1}{2}$ & $\frac{1}{2}$ & $\frac{1}{2}$ & $1$ & $1$ &
$\frac{1}{2}$\\ \rule[-2mm]{0mm}{5mm}  
$5$ & $\frac{1}{2}$ & $\frac{1}{2}$ & $\frac{1}{2}$ & $1$ & $1$ & $\frac{1}{2}$
\\ \rule[-2mm]{0mm}{5mm}  
$6$ & $1$ & $1$ & $1$ & $2$ & $2$ & $2$
\end{tabular} 
\end{center} 
\end{table}  

Note that the $g_{mm'}$ include some zeros and some negative values. To make
sense of these peculiarities we rewrite each of the six binomial factors of
(\ref{eq:48aw6}) in the form
\begin{equation}
  \label{eq:53jlj}
  \left( \begin{tabular}{c}
$B_m+(1-g_{mm})X_m-Y_m$ \\ $X_m$
\end{tabular} \right),
\end{equation}
where
\begin{equation}
  \label{eq:54jlj}
  B_m\doteq  A_m+g_{mm},\quad Y_m\doteq \sum_{m'\neq m}g_{mm'}X_{m'}+1.
\end{equation}
The maximum capacity for particles of species $m$,
\begin{equation}
  \label{eq:55jlj}
  X_m^\mathrm{max}=\frac{B_m-Y_m}{g_{mm}}.
\end{equation}
is thus primarily dictated by the diagonal coefficient $g_{mm}$, but is also
influenced by the off-diagonal coefficients $g_{mm'}$ via $Y_m$.

If one of the off-diagonal coefficients is zero, $g_{mm'}=0$ for $m'\neq m$, it
merely means that the presence of particles of species $m'$ has no effect on
the capacity for particles of species $m$. If one of the diagonal coefficients
vanishes, $g_{mm}=0$, then (\ref{eq:53jlj}) does no longer limit the capacity
for particles of species $m$. This can either mean that there is no limit (as
is the case for bosons) or it can mean (as is the case here for
$m=3$) that a limit is implied by a different rule associated with
(\ref{eq:48aw6}).

The existence of negative off-diagonal coefficients $g_{mm'}$ as found in
Table~\ref{tab:gmmps1w6} for $m=3$ and $m'=4,5,6$ has the consequence that
particles from species $m'$ contribute negatively to $Y_{m}$. Adding particles
of species $m'$ increases the capacity of the system for particles of species
$m$. This is indeed to be expected because the latter can only exist inside the
former. In this instance, the Pauli \emph{exclusion} principle turns into what
might be called an \emph{accommodation} principle.


\subsection{Statistical mechanics of soliton pairs}\label{sec:stamesopaga}

Carrying out the statistical mechanical analysis of $\mathcal{H}_1$ as a
gas of soliton pairs starts with solving six coupled nonlinear equations of the
type (\ref{eq:54bb}) with $\mu=0$, the $\epsilon_m$ from Table~\ref{tab:solpqp1}, and the
$g_{mm'}$ from Table~\ref{tab:gmmps1w6}. Symmetry implies~ $w_1=w_2,~ w_4=w_5$.
The remaining four equations in exponentiated form (with $K\doteq -J/k_BT$),
\begin{equation}
  \label{eq:82jlj}
  e^{-2K}=\frac{w_1w_4w_6}{(1\!+\!w_4)(1\!+\!w_6)},\quad
  e^{-K}=\frac{(1\!+\!w_3)w_1w_4w_6}{(1\!+\!w_1)(1\!+\!w_4)(1\!+\!w_6)}, 
\end{equation}
\begin{equation}
  \label{eq:84jlj}
\fl   e^{-2K}=\frac{w_1(1+w_3)w_4^2w_6^2}{(1+w_1)w_3(1+w_4)(1+w_6)^2}, \quad
   e^{-2K}=\frac{w_1(1+w_3)w_4w_6^2}{(1+w_1)w_3(1+w_4)(1+w_6)},
\end{equation}
can be simplified into
\begin{equation}
  \label{eq:86jlj}
  \hspace*{-15mm}\frac{1+w_3}{1+w_1}=e^K,\quad \frac{w_3}{w_6}=e^K,\quad w_4=1+w_6,\quad
  \frac{2+w_6}{w_1w_6}=e^{2K},
\end{equation}
and reduced to a quadratic equation for $w_3$ with the (physically significant) solution
\begin{equation}
  \label{eq:87jlj}
  w_3=\cosh K -\frac{1}{2} +\sqrt{\left(\cosh K-\frac{1}{2}\right)^2+2}.
\end{equation}
The grand partition function (\ref{eq:71bb}) with the (asymptotic) $A_m$ from
Table~\ref{tab:solpqp1} and the solutions (\ref{eq:86jlj}), (\ref{eq:87jlj})
becomes
\begin{equation}
  \label{eq:88jlj}
 \fl Z =
   \left[\frac{(1+w_1)^2}{w_1^2}\frac{(1+w_4)^2}{w_4^2}
\frac{(1+w_6)^2}{w_6^2}\right]^{N/2} 
   = \left[(1+w_3)e^{K}\right]^N=Z_Ne^{KN},
\end{equation}
in agreement with the transfer-matrix result for the canonical partition
function $Z_N$ \cite{STK67}, where the factor $e^{KN}$ again originates from a
shift in energy scale.

\begin{figure}[htb]
  \centering
  \includegraphics[width=70mm,angle=-90]{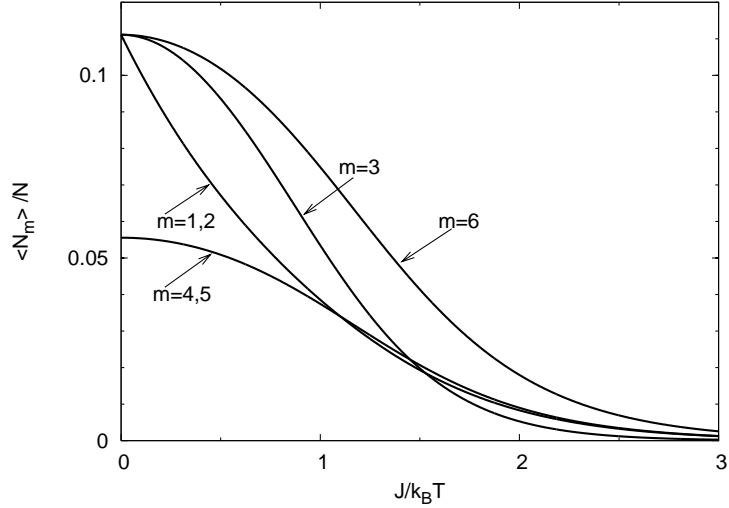}  
  \caption{Average numbers $\langle N_m\rangle/N$ of soliton-pair
    particles (per site) versus inverse temperature $J/k_BT$ of $\mathcal{H}_1$.}
  \label{fig:qpis16a}
\end{figure}

For the average numbers of soliton pairs, $\langle N_m\rangle$, we must solve
six linear equations of the type (\ref{eq:16appb}) with the $w_m$ from
(\ref{eq:86jlj}), (\ref{eq:87jlj}). Symmetry dictates that $\langle
N_1\rangle=\langle N_2\rangle$ and $\langle N_4\rangle=\langle N_5\rangle$. The
solution reads
\begin{eqnarray}
  \label{eq:105}
\langle N_1\rangle&=\langle N_2\rangle =\frac{N}{2}
\frac{w_3(w_3^2+2e^{K})}{(w_3+1)(w_3^2e^{-K}+4w_3+2-2e^{K})},
  \\ \label{eq:106}
  \langle N_3\rangle &= N\frac{2(w_3+1-e^K)}{(w_3+1)(w_3^2e^{-K}+4w_3+2-2e^{K})},
  \\ \label{eq:107}
  \langle N_4\rangle &=\langle N_5\rangle=\frac{1}{4}w_3\langle
  N_3\rangle,\quad 
\langle N_6\rangle=\frac{1}{2}w_3\langle N_3\rangle.
\end{eqnarray}
The relation $\langle N_6\rangle=2\langle N_4\rangle=2\langle N_5\rangle$ may
be anticipated on the basis of the motif shown in Table~\ref{tab:solpqp1}. The
reduced averages $\langle N_{m}\rangle/N$ are plotted versus $J/k_{B}T$ in
Fig.~\ref{fig:qpis16a}.

Increasing the temperature from $T=0$ results in a gradual increase of average
particle numbers from all species. Notice that the spacer particles $r_0$
($m=3$), which can only exist inside particles of species $q_+,q_0,q_-$
($m=4,5,6$) are slowest to appear in significant numbers as would be
expected. 

For the generalization of these results to $h\neq0$ we must add the Zeeman
contribution to the energies $\epsilon_{m}$ listed in Table~\ref{tab:solpqp1}.
The statistical mechanical analysis of soliton pairs as demonstrated here is by
no means limited to the Ising Hamiltonian (\ref{eq:10}). We can freeze out some
of the particle species by making their activation energies infinitely large,
$\epsilon_{m}\to\infty$. This has the consequence that $w_{m}\to0$ and $\langle
N_{m}\rangle\to0$. For the remaining active particles we can assign arbitrary
values $\epsilon_{m}$ for their energies. For example, if we freeze out all
particles except those of species $m=1,2$ then the results of
Sec.~\ref{sec:isoh} for $\mathcal{H}_{1/2}$ are, effectively, recovered.

Among the issues that must be heeded in generalizations of the calculations
reported here to models with arbitrary particle energies are the following: (i)
The particle $r_{0}$ can only exist inside a particle $q_{0}$, $q_{+}$, or
$q_{-}$. Therefore freezing out the latter three will freeze out the former
even if its energy remains finite.  (ii) The particle $q_{0}$ comes in two
parity-violating versions. In the context of $\mathcal{H}_1$ or other models
where the two configurations $\uparrow\circ\downarrow$ and
$\downarrow\circ\uparrow$ have the same energy, they can be treated as
identical particles. However, in situations where the two configurations have
to be assigned different energies we must treat them as belonging to different
species and determine their statistical interaction with each other and with
all the other particle species. (iii) A spin interaction beyond nearest neighbors
added to $\mathcal{H}_1$ will, in general, produce a coupling between the
particles listed in Table~\ref{tab:solpqp1}.


%
\section{Domains and nested domains}\label{sec:stri}
%
Here we consider $\mathcal{H}_s$ for $s=1/2,1$, $J<0$, and $h\leq 0$. In the
following we describe how the entire spectrum is systematically generated from
the ferromagnetic ground state $|\uparrow\uparrow\cdots\uparrow\rangle$ by
domains $(s=1/2)$ or nested domains $(s=1)$ of flipped spins. These domains are
independent particles subject to a statistical interaction. The thermodynamic
analysis of domains is then carried out for the $s=1/2$ case at $h=0$.

\subsection{Combinatorics of domains}\label{sec:comdom}

In the notation used here, $\{ |\sigma_1\cdots\sigma_N\rangle \}^{r}_m$
represents the set of $m$ product vectors with $r$ flipped spins that are
generated from $|\sigma_1\cdots\sigma_N\rangle$ via translations.  The $2^N=16$
states for $N=4$ in this representation are
\begin{eqnarray}
  \label{eq:7}
 &&  \{|\uparrow\uparrow\uparrow\uparrow\rangle\}^{0}_1,\quad
 \{|\uparrow\uparrow\uparrow\downarrow\rangle\}^{1}_4,\quad
 \{|\uparrow\uparrow\downarrow\downarrow\rangle\}^{2}_4,\nonumber \\ 
&& \{|\uparrow\downarrow\uparrow\downarrow\rangle\}^{2}_2,\quad
\{|\uparrow\downarrow\downarrow\downarrow\rangle\}^{3}_4,\quad
\{|\downarrow\downarrow\downarrow\downarrow\rangle\}^{4}_1. 
\end{eqnarray}
The first among them is the (non-degenerate) ground state of
$\mathcal{H}_{1/2}$ with $J<0$ and $h<0$. Domains are strings of $\mu$
consecutive down-spins. In (\ref{eq:7}) the states in the second set contain
one 1-string, and the state in the fourth set two 1-strings. The states in the
third, fifth, and sixth set contain one string with $\mu=2,3,4$, respectively.
Each string of length $\mu$ contributes the amount $J+\mu h$ to the energy of the
state. An Ising chain of length $N$ can thus accommodate strings with
$\mu=1,\ldots,N-1$, which are treated here as distinct species of independent
particles. The lone state containing one string with $\mu=N$ is exceptional in
several respects, ignorable in macroscopic systems.

What is the number of product eigenstates that
contain a configuration $\{N_\mu\}$ of strings? Since there must be at least
one up-spin between successive strings, only those configurations can be
realized which satisfy the constraint
\begin{equation}
  \label{eq:55clj}
  \sum_{\mu=1}^{N-1} (\mu+1) N_\mu\leq N.
\end{equation}
We found that the number of states with given string
configuration is determined by the multiplicity expression
\begin{equation}
  \label{eq:57clj}
   W(\{N_\mu\}) = \frac{N}{N-r}\;\prod_{\mu=1}^{N-1}   
\left(
\begin{tabular}{c}
$d_\mu +N_\mu-1$ \\ $N_\mu$
\end{tabular}
\right),
\end{equation}
\begin{equation}
  \label{eq:58clj}
  d_\mu=N-\mu -\sum_{\mu'=1}^{N-1} g_{\mu\mu'}(N_{\mu'}-\delta_{\mu\mu'}),
\end{equation}
where
\begin{equation}
  \label{eq:59clj}
  g_{\mu\mu'}=\left\{
\begin{tabular}{ll}
$\mu'$, & $\mu<\mu'$, \\ $\mu'+1$, & $\mu\geq\mu'$
\end{tabular}
\right.,\qquad  r\doteq \sum_{\mu=1}^{N-1} \mu N_{\mu}.
\end{equation}


\subsection{Combinatorics of nested 
domains}\label{sec:comnesdom}

The concept of nested quasiparticles in lattice models is well-known in the
context of the Bethe ansatz as applied, for example, to the Hubbard model or to
integrable spin-1 models \cite{EFG+05, Suth75, Takh82, Babu82, Hald82a, Sogo84,
  FS98}. The nested particles in Ising product states have a particularly
simple structure.

In the context of $\mathcal{H}_1$ the nesting involves two shells. The
particles on the outer shell ($\mu$-strings) are structurally identical to the
strings of $\mathcal{H}_{1/2}$.
We start from the $\mu$-string vacuum,
$\{|\uparrow\uparrow\cdots\uparrow\rangle\}_1^{0}$, and generate a total of
$2^N$ product states composed of site variables $\uparrow$ and $\circ$.  On the
inner shell we take any $\mu$-string of the outer shell and use it as the
vacuum for $\nu$-strings. Hence a $\nu$-string is a sequence of $\nu$
successive $\downarrow$-sites embedded in a region of $\mu$ $\circ$-sites
between consecutive $\uparrow$-sites. Naturally, we must have $\nu\leq\mu$.
This prescription is illustrated in Table~\ref{tab:nest4} for $N=4$.
\begin{table}[htb]
  \caption{Nested-string interpretation of the product
    eigenstates for $N=4$ in $\mathcal{H}_1$. The $2^N$ states on the left
    represent the outer shell of the 
    nesting. Each $\mu$-string is underlined and serves as the vacuum for
    $\nu$-strings on the inner shell. From each of the $2^N$
    effective $\mathcal{H}_{1/2}$ product states on the left are thus generated
    one or   several $\mathcal{H}_1$ product states on the right for a total of
    $3^N$.} 
\label{tab:nest4} 
\begin{center}
\begin{tabular}{rl|lr} \hline \rule[-2mm]{0mm}{7mm}
1 & $\{|\uparrow\uparrow\uparrow\uparrow\rangle\}^{+4}_1$ & 
$\{|\uparrow\uparrow\uparrow\uparrow\rangle\}^{+4}_{1\times1}$,
 & 1 \\
\rule[-2mm]{0mm}{6mm}
4 & $\{|\uparrow\uparrow\uparrow\underline{\circ}\rangle\}^{+3}_4$ & 
$\{|\uparrow\uparrow\uparrow\circ\rangle\}^{+3}_{4\times1}$,
$\{|\uparrow\uparrow\uparrow\downarrow\rangle\}^{+2}_{4\times1}$,
 & 8 \\
\rule[-2mm]{0mm}{6mm}
4 & $\{|\uparrow\uparrow\underline{\circ\circ}\rangle\}^{+2}_4$ & 
$\{|\uparrow\uparrow\circ\circ\rangle\}^{+2}_{4\times1}$, 
$\{|\uparrow\uparrow\downarrow\circ\rangle\}^{+1}_{4\times2}$,
$\{|\uparrow\uparrow\downarrow\downarrow\rangle\}^{0}_{4\times1}$,
 &16  \\
\rule[-2mm]{0mm}{6mm}
4 & $\{|\uparrow\underline{\circ\circ\circ}\rangle\}^{+1}_4$ & 
 $\{|\uparrow\circ\circ\circ\rangle\}^{+1}_{4\times1}$, 
$\{|\uparrow\downarrow\circ\circ\rangle\}^{0}_{4\times3}$, 
$\{|\uparrow\downarrow\downarrow\circ\rangle\}^{-1}_{4\times3}$,
$\{|\uparrow\downarrow\downarrow\downarrow\rangle\}^{-2}_{4\times1}$  
& 32 \\
\rule[-2mm]{0mm}{7mm}
1 & $\{|\underline{\circ\circ\circ\circ}\rangle\}^{0}_1$ & 
 $\{|\circ\circ\circ\circ\rangle\}^{0}_{1\times1}$, 
$\{|\downarrow\circ\circ\circ\rangle\}^{-1}_{1\times4}$  \\ 
&&$\{|\downarrow\downarrow\circ\circ\rangle\}^{-2}_{1\times4}$, 
$\{|\downarrow\circ\downarrow\circ\rangle\}^{-2}_{1\times2}$,
$\{|\downarrow\downarrow\downarrow\circ\rangle\}^{-3}_{1\times4}$,
$\{|\downarrow\downarrow\downarrow\downarrow\rangle\}^{-4}_{1\times1}$,
 & 16 \\
\rule[-2mm]{0mm}{7mm}
2 & $\{|\uparrow\underline{\circ}\uparrow\underline{\circ}\rangle\}^{+2}_2$ & 
 $\{|\uparrow\circ\uparrow\circ\rangle\}^{+2}_{2\times1}$, 
$\{|\uparrow\downarrow\uparrow\circ\rangle\}^{+1}_{4\times1}$,
$\{|\uparrow\downarrow\uparrow\downarrow\rangle\}^{0}_{2\times1}$, 
& 8 \\ \hline
\rule[-2mm]{0mm}{7mm} 
16 & & & 81
\end{tabular}
\end{center}
\end{table}
The two-shell nesting of string particles leads to the multiplicity
expression
\begin{eqnarray}
  \label{eq:20plj1}
  W\left(\{N_\mu\},\{N_\nu^{(\mu)}\}\right) &=& \frac{N}{N-r}\;\prod_\mu   
\left(
\begin{tabular}{c}
$d_\mu +N_\mu-1$ \\ $N_\mu$ 
\end{tabular} \right)
\nonumber \\
&~~~~\times& \frac{\mu}{\mu-r_\mu}\;\prod_\nu   
\left(
\begin{tabular}{c}
$d_\nu^{(\mu)} +N_\nu^{(\mu)}-1$ \\ $N_\nu^{(\mu)}$
\end{tabular}
\right)
\end{eqnarray}
with $d_{\mu}$ from (\ref{eq:58clj}), $g_{\mu\mu'}$, $r$ from (\ref{eq:59clj}),
and  
\begin{equation}
  \label{eq:22plj1}
  d_\nu^{(\mu)}=\mu -\nu  -\sum_{\nu'} 
g_{\nu\nu'}(N_{\nu'}^{(\mu)}-\delta_{\nu\nu'}),\qquad
r_\mu=\sum_\nu\nu N_\nu^{(\mu)}.
\end{equation}
As in previous applications, there are instances (ignorable for macroscopic
systems) where expression (\ref{eq:20plj1}) is inapplicable.

\subsection{Statistical mechanics of 
domains}\label{sec:stamemu}

Returning to $\mathcal{H}_{1/2}$ with $J<0$ and setting $h=0$, we now derive
the exact thermodynamics of a system of strings via the method outlined in
Sec.~\ref{sec:statint}. It is evident from Wu's derivation \cite{Wu94} of
Eqs.~(\ref{eq:54bb}) that their applicability in the present context is limited
to situations where the system has a large capacity for strings of all sizes
that are permitted.  To circumnavigate this restriction we introduce a limit
on the length of allowed strings, $\mu\leq M\ll N$. The thermodynamic limit of
$\mathcal{H}_{1/2}$ requires that we set $N\to\infty$ before setting
$M\to\infty$.

With the specifications regarding statistical interaction of strings from
Sec.~\ref{sec:comdom} we write for the grand potential the expression
\begin{equation}
  \label{eq:8}
  \Omega_{M}(K)=-k_BT\sum_{\mu=1}^MA_\mu\ln\left(\frac{w_\mu+1}{w_\mu}\right),\quad 
A_\mu=N-\mu,
\end{equation}
where the $w_\mu$ satisfy
\begin{equation}
  \label{eq:9}
  4K=\ln(w_\mu+1) -\sum_{\mu'=1}^M
  g_{\mu'\mu}\,\ln\frac{w_{\mu'}+1}{w_{\mu'}},\quad K=\frac{|J|}{4k_BT}. 
\end{equation}
The transformation of variable, $\xi_\mu\doteq\ln(w_\mu+1)$, turns Eqs.~(\ref{eq:8}) and
(\ref{eq:9}) into 
\begin{equation}
  \label{eq:31}
  \Omega_M(K)= \frac{|J|}{4K}\sum_{\mu=1}^M(N-\mu)\ln\left(1-e^{-\xi_\mu}\right),
\end{equation}
\begin{equation}
  \label{eq:32}
  \xi_\mu=4K-\mu\sum_{\mu'=1}^M\ln\left(1-e^{-\xi_{\mu'}}\right)
  -\sum_{\mu'=\mu}^M\ln\left(1-e^{-\xi_{\mu'}}\right). 
\end{equation}
Introducing the quantity
\begin{equation}
  \label{eq:33}
  \Phi_\mu\doteq-\frac{1}{4K}\sum_{\mu'=1}^\mu \ln\left(1-e^{-\xi_{\mu'}}\right)
\end{equation}
we rewrite (\ref{eq:32}) in the form
\begin{equation}
  \label{eq:34}
  \xi_\mu=4K\left[1+(\mu+1)\Phi_M-\Phi_{\mu-1}\right].
\end{equation}
This sets the stage for determining $\Phi_M$ via a recursive scheme:
\begin{equation*}
  \Phi_1=-\frac{1}{4K}\ln\left(1-q^{1+2\Phi_M}\right),\quad q\doteq e^{-4K},
\end{equation*}
\begin{equation*}
  \Phi_2=\Phi_1-\frac{1}{4K}\ln\left(1-\exp\left(-4K(1+3\Phi_M)
    -\ln\left(1-q^{1+2\Phi_M}\right)\right)\right), 
\end{equation*}
leading to
\begin{equation}
  \label{eq:38}
\Phi_\mu = -\frac{1}{4K}\ln\left(1-q^{1 +2\Phi_M}\frac{1
-q^{\mu \Phi_M}}{1-q^{\Phi_M}}\right).
\end{equation}
Setting $\mu=M$ we arrive at a polynomial equation for
$q^{\Phi_M}$:
\begin{equation}
  \label{eq:39}
  q\,q^{(M+2)\Phi_M}+(1-q)q^{2\Phi_M}-2q^{\Phi_M}+1=0.
\end{equation}
The solution of (\ref{eq:39}) substituted into (\ref{eq:31}) via (\ref{eq:38})
and (\ref{eq:34}) determines the grand potential of a system of strings with
maximum length $M$ in a chain of $N$ sites with $M\ll N$. Taking the limit
$N\to\infty$ while keeping $M$ finite we have
\begin{equation}
  \label{eq:40}
 \omega_M(K)\doteq\! \lim_{N\to\infty}\frac{1}{N}\Omega_M(K)=
  \frac{|J|}{4K}\sum_{\mu=1}^{M}\ln\left(1\!-\!e^{-\xi_\mu}\right) =-|J|\Phi_M.
\end{equation}
If we now take the limit $M\to\infty$, the first term in Eq.~(\ref{eq:39})
vanishes, and the solution,
\begin{equation}
  \label{eq:45}
  q^{\Phi_\infty}=(1+\sqrt{q})^{-1},
\end{equation}
substituted into (\ref{eq:40}), yields
\begin{equation}
  \label{eq:41}
  \omega_\infty(K)=-\frac{|J|}{4K}\ln\left(1+e^{-2K}\right),
\end{equation}
which is indeed the exact result for $\mathcal{H}_{1/2}$ with $h=0$, $J<0$
and the string vacuum at the origin of the energy scale.

The statistical mechanics of a system of domains with maximum length $M$ on a
lattice of $N$ sites may very well be of interest in a number of contexts
outside magnetism. We have reduced the problem to solving a polynomial equation
of degree $M+1$. Consider the entropy per site of strings with $\mu\leq M$ on
an infinite lattice, $s_{M}(K)\doteq \lim_{N\to\infty} S_{M}(K)/N$, inferred
from (\ref{eq:40}). Compact analytic solutions are readily
calculated for $M=1$ (one-strings only) and $M=\infty$ (all strings allowed):
\begin{equation}
  \label{eq:42}
  \hspace*{-1cm}
\frac{s_{1}(K)}{k_B}= \ln\left(\frac{\sqrt{1+4e^{-4K}}+1}{2}\right)
  +\frac{8Ke^{-4K}}{1+4e^{-4K}+\sqrt{1+4e^{-4K}}}, 
\end{equation}
\begin{equation}
  \label{eq:43}
 \hspace*{-1cm}
  \frac{s_{\infty}(K)}{k_B}= \ln\left(1+e^{-2K}\right) +\frac{2K}{e^{2K}+1}.
\end{equation}

\begin{figure}[tb]
  \centering
  \includegraphics[width=67mm,angle=-90]{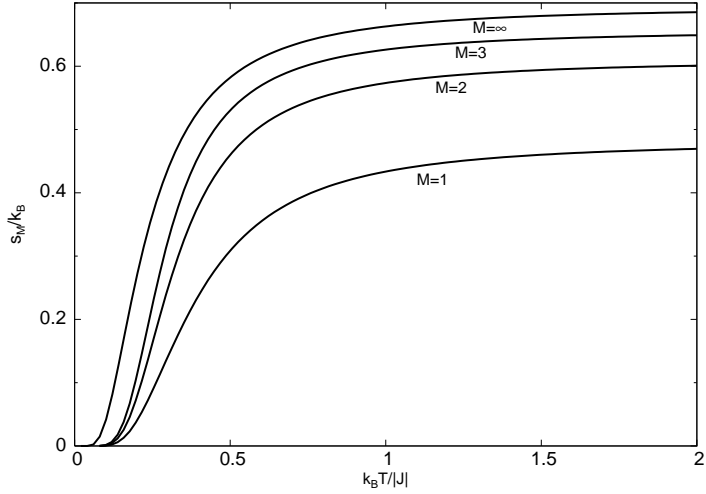}  
  \caption{Entropy per site for $N\to\infty$ versus reduced temperature
    of a system of domains with maximum length $M$. The case $M=\infty$
    represents $\mathcal{H}_{1/2}$.}
  \label{fig:entropy}
\end{figure}

Entropy curves for several $M$ are shown in Fig.~\ref{fig:entropy}. As we relax
the restriction on the length of permissible domains, the entropy at any given
nonzero temperature becomes larger. The relative contribution of longer domains
is larger at low $T$ than at high $T$. All domains have the same energy. With
$T$ increasing, the longer domains tend to be crowded out by the shorter ones.
As the restriction on length is lifted altogether, the Ising result
$(M=\infty)$ is approached from below.

The same type of analysis is applicable to any model with spin-1/2 Ising
product eigenstates and with arbitrary energy values $\epsilon_{m}$,
$\mu=1,2,\ldots,M$ assigned to the domains. The left-hand side of
Eq.~(\ref{eq:9}) must then be replaced by $4K_{\mu}$,
$K_{\mu}=\epsilon_{\mu}/4k_{B}T$. For $\mathcal{H}_{1/2}$ at $h\neq 0$ we must
use $\epsilon_{\mu}=J+\mu h$.

\subsection{Distribution of domains}\label{sec:disdom}

What is the relative frequency of occurrence of domains of size $\mu$ for given
maximum size $M$ at temperature $T$ in an infinite chain? To answer this
question we adapt Wu's linear equations (\ref{eq:16appb}) to the situation at
hand:
\begin{equation}
  \label{eq:44}
  w_\mu\langle n_\mu\rangle +\sum_{\mu'=1}^M\mu'\langle n_{\mu'}\rangle + \sum_{\mu'=1}^\mu\langle
  n_{\mu'}\rangle=1,\quad \mu=1,\ldots,M,
\end{equation}
where $n_\mu\doteq N_\mu/N$ and where we have ignored a contribution of
O$(\mu/N)$ to the right-hand side by effectively taking the limit $N\to\infty$,
while keeping $M$ finite. The quantities $w_\mu=e^{\xi_\mu}-1$ are known from
the solution of (\ref{eq:39}) via (\ref{eq:38}) and (\ref{eq:34}).

Here we carry out the calculation for the case $M\to\infty$. The solution
(\ref{eq:45}) substituted into (\ref{eq:38}) yields
\begin{equation}
  \label{eq:46}
  q^{\Phi_{\mu}} =\frac{1}{1+\sqrt{q}}
  \left(1+ \frac{\sqrt{q}}{(1+\sqrt{q})^{\mu}}\right)\qquad (M=\infty),
\end{equation}
which, upon substitution in (\ref{eq:34}), produces the $w_{\mu}$ needed in
(\ref{eq:44}):
\begin{equation}
  \label{eq:47}
  w_{\mu}= \frac{1}{\sqrt{q}} + \frac{(1+\sqrt{q})^{\mu}}{q}\qquad (M=\infty).
\end{equation}
Now we rewrite Eqs.~(\ref{eq:44}) in the form
\begin{equation}
  \label{eq:48}
  w_\mu\langle n_\mu\rangle + \sum_{\nu=1}^\mu\langle n_\nu\rangle
=\zeta,\quad \mu=1,2,\ldots 
\end{equation}
where the quantity
\begin{equation}
  \label{eq:49}
  \zeta\doteq 1-\sum_{\nu=1}^\infty\nu\langle n_\nu\rangle
\end{equation}
can be treated as a constant to be determined self-consistently at the end.
The solution of Eqs. (\ref{eq:48}), obtained by induction, is 
\begin{equation}
  \label{eq:51}
  \langle n_\mu\rangle= \frac{P_{\mu}}{w_{\mu}}
\left(1+\sum_{\nu=1}^{\infty}\frac{\nu}{w_{\nu}}P_{\nu}\right)^{-1},\quad 
P_{\mu}\doteq\prod_{\nu=1}^{\mu}\frac{w_{\nu}}{w_{\nu}+1},
\end{equation}
and, after normalization, 
\begin{equation}
  \label{eq:53}\hspace*{-10mm}
  \langle \hat{n}_{\mu}\rangle \doteq
\langle n_\mu\rangle \left(\sum_{\nu=1}^{\infty}\langle n_\nu\rangle\right)^{-1}
 = \frac{P_{\mu}}{w_{\mu}}\left(\sum_{\nu=1}^{\infty}
\frac{P_{\nu}}{w_{\nu}}\right)^{-1} =\frac{P_{\mu}}{w_{\mu}(1-P_{\infty})},   
\end{equation}
From (\ref{eq:33}) and (\ref{eq:51}) we infer that $P_{\mu}=q^{\Phi_{\mu}}$ a
quantity evaluated in (\ref{eq:46}). The assembly of the ingredients
(\ref{eq:46}), (\ref{eq:47}), and (\ref{eq:45}) to expression (\ref{eq:53})
produces the following explicit result for the distribution of lengths $\mu$ of
string particles in $\mathcal{H}_{1/2}$ at temperature $T$ and zero magnetic
field:
\begin{equation}
  \label{eq:55}
  \langle \hat{n}_{\mu}\rangle= \frac{\sqrt{q}}{(1+\sqrt{q})^{\mu}}
  =\frac{e^{-2K}}{(1+e^{-2K})^{\mu}},\quad \mu=1,2,\ldots 
\end{equation}
This is a realization of Pascal's distribution,
$P(\mu)=\gamma(1-\gamma)^{\mu-1}$, if we set $\gamma=e^{-K}/(e^{K}+e^{-K})$.
This result was previously derived by Denisov and H\"anggi \cite{DH05} using a
very different method in their study of finite Ising chains with open
boundaries.  This distribution indeed favors short strings in the crowded
conditions at high $T$, in agreement with observations made in our discussion
of the entropy curves (Fig.~\ref{fig:entropy}). At low $T$ the distribution is
flat, consistent with the fact that all strings have the same energy. With some
additional effort our solution can be generalized to finite $M$, and to models
with arbitrary values for the energies $\epsilon_{\mu}$ of domains of size
$\mu$.


%
\section{Conclusion}\label{sec:outlook}
%

We have demonstrated that the conceptual framework of statistical interaction
between quasiparticles in many-body systems \cite{Hald91a,Wu94} leads to
significant new insights into the statistical mechanics of Ising chains and
related models with spin-1/2 or spin-1 product eigenstates on a one-dimensional
lattice. We have identified, in particular, the nature of complementary sets of
independent particles on the basis of which the spectrum of Ising chains with
$s=1/2$ and $s=1$ can be generated systematically from either the ferro- or
antiferromagnetic ground state.

The N{\'e}el state is the vacuum for solitonic particles. In the $s=1/2$ case
the solitons themselves are the independent particles. They are
antiferromagnetic domain walls, confined to single bonds, with spin $\pm1/2$
and semionic statistical interaction. In the $s=1$ case the solitons are merely
building blocks of particles. All independent particles are soliton pairs. The
paired solitons may be on the same bond or on bonds any number of lattice units
apart. We have carried out the exact statistical mechanical analysis of
solitons (two species) for $s=1/2$ and of soliton pairs (six species) for
$s=1$.

The state with all spins up is the vacuum for string particles. In the
$s=1/2$ case the independent particles are domains of overturned spins and in
the $s=1$ case they are nested domains, i.e. domains inside domains of halfway
overturned spins. By working out their exact statistical interaction we have
set the stage for the statistical mechanical analysis of domains and nested
domains. We have carried out that analysis for the $s=1/2$ case and established
contact with previous work based on different methods \cite{DH05}.

The work presented here opens up numerous opportunities for extensions and
comparisons including the following. 
(i) The methodology developed in Secs.~\ref{sec:isoh} and \ref{sec:isone} for the
identification and specification of independent solitonic particles looks
promising for applications to Ising chains with $s>1$ and to Ising
ladders. Preliminary results for $\mathcal{H}_{3/2}$, for example, indicate that
the independent solitonic particles contain at least two and no more than
six solitons. This again includes particles confined to one bond and particles
spread across many bonds with more than one species of spacer particles acting
as deconfinement agents. 
(ii) A question of considerable interest is how the methodology developed here
can be generalized to situations with Ising interactions beyond nearest
neighbors, which, in general, leads to a coupling between solitonic particles
and between string particles. 
(iii) There exist integrable spin chain models with a parametric Ising limit.
Consequently, the solitonic particles analyzed here must exist in some variant
form away from the Ising limit of those models. One such link (to the spinons
of the $s=1/2$ $XXZ$ model) is outlined in \ref{sec:nestri}.  Corresponding
links are bound to exist between the string and nested-string particles of
Ising chains and the string solutions of the Bethe ansatz applied to integrable
spin models with $s=1/2$ \cite{Gaud71, Taka91, JM72, John74, JB80, Taka99} and
$s=1$ \cite{Suth75, Takh82, Babu82, Hald82a, Sogo84, FS98} near their Ising
limits.


\appendix

%
\section{Solitons versus spinons}\label{sec:nestri}
%
The ground state of the $s=1/2$ $XXZ$ model,
\begin{equation}
  \label{eq:18}
  \mathcal{H}_{XXZ}=
\sum_{n=1}^N\left[J_\perp(S_n^xS_{n+1}^x+S_n^yS_{n+1}^y) +J_zS_n^zS_{n+1}^z\right],
\end{equation}
at $J_\perp, J_z \geq0$ for even $N$ is non-degenerate except in the Ising limit
$(J_\perp=0)$. The finite-size gap is of O$(N^{-1})$ in the planar regime
$(J_\perp>J_z)$ and of O$(e^{-N})$ in the axial regime $(J_\perp<J_z)$. The lowest
energy level in both regimes has been identified as the (unique) vacuum of
spinons \cite{Taka99}. The two lowest levels, again in both regimes, can be identified as the
(twofold) vacuum of solitons.\footnote{The names attributed to quasiparticles
  in quantum spin chains vary among authors. Our usage is common but not
  universal.}

Spinons and solitons have similar but not identical semionic exclusion
statistics. The similarities and differences are encoded in the multiplicity
expressions. Equations (\ref{eq:7clj})-(\ref{eq:8clj}) for solitons are to be
compared with
\begin{equation}
  \label{eq:7spclj}
  W_S(N_+,N_-) = \prod_{\sigma=\pm} \left(
\begin{tabular}{c}
$d_\sigma +N_\sigma -1$ \\ $N_\sigma$
\end{tabular}
\right),
\end{equation}
\begin{equation}
  \label{eq:8spclj}
  d_\sigma=\frac{1}{2}(N+1) 
-\frac{1}{2}\sum_{\sigma'}(N_{\sigma'}-\delta_{\sigma\sigma'})
\end{equation}
for spinons \cite{Hald91a}.  Away from the Ising limit, solitons (and spinons)
are dispersive and scatter off each other elastically. Both kinds of
particles are most conveniently identified by their momentum quantum
numbers. Every $XXZ$ eigenstate has a unique spinon composition and a unique
soliton composition. The relation between the spinon composition and the
soliton composition is most transparent in the $XX$ limit $(J_z=0)$.

\begin{figure}[b]
  \centering
  \includegraphics[width=85mm]{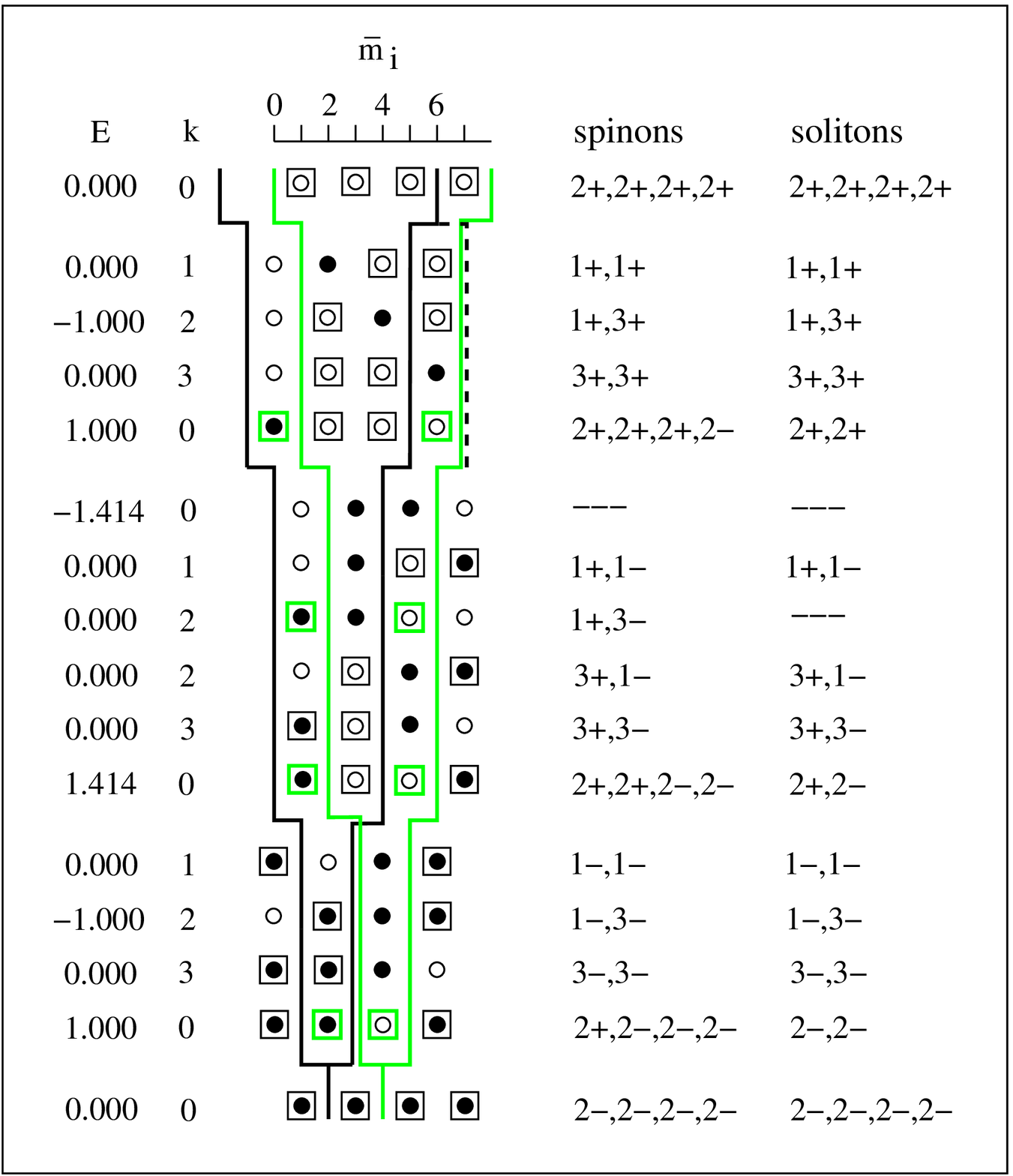}  
  \caption{Fermion configurations of all eigenstates for $N=4$ of the $XX$
    model $(J_z=0)$. Fermionic particles (holes) are denoted by full (open)
    circles.  Spinons with spin up (down) are denoted by (black or gray)
    squares around open (full) circles. Solitons with spin up (down) are
    denoted by black squares around open (full) circles. The fermion momenta
    $\bar{m}_i$ (in units of $\pi/N$) can be read off the diagram. The spinon
    orbital momenta $m_i$ (also in units of $\pi/N$) and the spinon spins
    $\sigma_i$ are given explicitly and can be inferred from the fermion
    configuration as explained in the text. Also given are the wave number $k$
    (in units of $2\pi/N$) and the energy $E$ (in units of $J_\perp$) of each
    eigenstate.}
  \label{fig:qpsp25a}
\end{figure}

In Ref. \cite{AKMW06} a motif was developed that relates the configuration of
(free) Jordan-Wigner fermions with the configuration of spinons. This motif is
reproduced in Fig.~\ref{fig:qpsp25a} for $N=4$ (16 eigenstates) and amended to
also show the soliton configuration.
The allowed fermion momenta (in units of $\pi/N$) are
\begin{equation}
  \label{eq:13akmw}
\bar{m}_i\in \left\{ 
\begin{tabular}{ll}
${\displaystyle \{1,3,\ldots,2N-1\}}$ & for even $N_F$ \\
${\displaystyle \{0,2,\ldots,2N-2\}}$ & for odd $N_F$
\end{tabular} \right.
\end{equation}
and the allowed spinon orbital momenta (in units of $\pi/N$) are
\begin{equation}
  \label{eq:lj66akmw}
m_i = \frac{N_S}{2}, \frac{N_S}{2}+2, \ldots,
  N-\frac{N_S}{2},  
\end{equation}
where $N_F$ is the number of fermions and $N_S=N_+ +N_-$ the number of spinons
in any given $XX$ eigenstate.

The exact spinon configuration is encoded in the fermion configuration as
described in the following: (i) Consider the the gray fork as dividing the
fermion momentum space into two domains, the inside and the outside. The
outside domain wraps around at the extremes ($\bar{m}_i=N\,\mathrm{mod}\,N=0$).
(ii) Every fermionic hole (open circle) inside represents a spin-up spinon
(square surrounding open circle) and every fermionic particle (full circle)
outside represents a spin-down spinon (square surrounding full circle).  (iii)
Any number of adjacent spinons in the representation of Fig.~\ref{fig:qpsp25a}
are in the same orbital. Two spin-up (spin-down) spinons that are
separated by $\ell$ consecutive fermionic particles (holes) have quantum
numbers separated by $2\ell$.  (iv) The spinon orbital momenta are sorted in
increasing order from the right-hand prong of the gray fork toward the left
across the inside domain and toward the right with wrap-around through the
outside domain.

For the determination of the soliton content of any $XX$ eigenstate we must
also consider the black fork in Fig.~\ref{fig:qpsp25a}, which is shifted two
units to the left. If there is any spin-up/spin-down spinon pair as identified
by the rules pertaining to the gray fork that does not also satisfy the same
rules with respect to the black fork, then it is omitted from the list of
solitons. All such spinons are identified by gray squares in
Fig.~\ref{fig:qpsp25a}.  The spinon vacuum is just one of two soliton vacua,
the other soliton vacuum being a two-spinon state. The two soliton vacua have
wave numbers differing by $\pi$. In the Ising limit of the $XXZ$ model they
correspond to the symmetric and antisymmetric combinations of the two product
N{\'e}el states.

%
\ack
%
Financial support from the DFG Schwerpunkt \textit{Kollektive Quantenzust{\"a}nde
  in elektronischen 1D {\"U}bergangsmetallverbindungen} (for M.K.)  is gratefully
acknowledged.

%
\section*{References} 
%

\end{document}